\documentstyle[preprint,eqsecnum,aps]{revtex}
\tightenlines

\def\be{\begin{eqnarray}}
\def\ee{\end{eqnarray}}
\def\G{\Gamma}
\def\D{\Delta}
\begin{document}
\draft
\preprint{ULB-TH-99/27}
\title{Refining the anomaly consistency condition\\
}
\author{Glenn Barnich\cite{byline}}
\address{
Physique Th\'eorique et Math\'ematique, Universit\'e Libre de
Bruxelles,\\
Campus Plaine C.P. 231, B--1050 Bruxelles, Belgium
}
\maketitle
\begin{abstract}
In the extended antifield formalism, 
a quantum BRST differential for anomalous gauge theories is
constructed. Local BRST cohomological classes
are characterized, besides the form degree and the ghost number, 
by the length of their
descents and of their lifts, and this both in the standard and the extended 
antifield formalism. It is shown that during the BRST invariant
renormalization of a local BRST cohomological class, the  
anomaly that can appear is constrained to be a local BRST cohomological
class with a shorter descent and a longer lift than the given class.
As an application of both results, a simple approach to 
the Adler-Bardeen theorem for the non abelian gauge anomaly is proposed. 
It applies independently of the gauge fixing, of power counting restrictions 
and does not rely on the use of the Callan-Symanzik equation.  
\end{abstract}


\section{Introduction}
\label{sec:level1}

Upon quantization, symmetries of the classical action can be affected
by anomalies. These anomalies have been shown to satisfy
important consistency conditions following from the algebra
of the symmetries \cite{Wess:1971cm}. 
The consistency conditions have been elegantly reformulated as a
cohomological problem, first in the case of the gauge anomaly in Yang-Mills
theories\cite{Becchi:1974xu,Becchi:1974md,Becchi:1975nq} 
and then also in the case of global symmetries with a closed
algebra \cite{Voronov:1981du,Becchi:1988vc}. 

In the case of gauge theories, the 
formalism has been generalized some time ago 
\cite{Batalin:1981jr,Batalin:1983jr} to cover
generic gauge theories, and the expression of the anomaly consistency
in that context has been discussed \cite{Howe:1990pz,Troost:1990cu}. 
Only recently, the corresponding generalization 
in the case of global symmetries with a generic algebra has been achieved
\cite{Brandt:1998cz}.  

In the same way,
one can study the constraints on the
anomalies to an invariant renormalization of symmetric integrated and non
integrated operators \cite{Dixon:1980zp}. 
In the cohomological reformulation, symmetric
integrated or non integrated operators correspond to BRST cohomological
classes in ghost number $0$ in the space of local functionals,
respectively in the space of local functions. The
non trivial anomalies that can appear can be
shown to belong to the corresponding BRST cohomological classes in 
ghost number $1$. 
 
The computation in the space of local functionals 
of the cohomology of the standard BRST differential $s$ with antifields 
can be reduced to the computation of
a relative cohomological group in the space of local $n$-forms by
introducing the space-time exterior derivative $d$. It is then related to
the cohomology of $s$ in the space of form 
valued local functions through descent equations
\cite{Stora:1976kd,Stora:1983ct,Zumino:1983ew,Zumino:1984rz}. The same 
holds for the BRST differential $\bar s$ associated to the 
extended antifield formalism. 

More generally, we will call the relative cohomological groups
$H^{g,p}(s|d)$ and $H^{g,p}(\bar s|d)$ in ghost number $g$ and 
form degree $p$ local BRST
cohomological groups.
A more detailed analysis 
of the descent equations \cite{Dubois-Violette:1985jb} shows 
that these groups
are characterized by two integers, the length $d$ of their descents and the 
length $l$ of their lifts. 

The purpose of this article is twofold. First, it is shown 
that in the extended antifield
formalism, a quantum BRST differential can be constructed even for
anomalous theories, because the quantum version in terms of the
effective action of the extended master
equation can be written as a functional differential equation. Then it 
is shown that both in the extended antifield formalism and in the
standard antifield formalism, the anomaly appearing 
in the renormalization of a local BRST cohomological class with a
descent of length $d$ and a lift of length $l$ 
are characterized by a descent which is shorter or equal to $d$ and a
lift which is longer or equal to $l$. 

As an application, it is shown that the anomalous master equation for
Yang-Mills theories discussed in 
\cite{Costa:1977pd,Aoyama:1981yw,Tonin:1992wf} can be viewed as a
particular case of the anomalous master equation of the extended
antifield formalism. Then, we discuss how the Adler-Bardeen theorem
for the non abelian gauge anomaly
\cite{Bardeen,Costa:1977pd,Bandelloni:1980sp,Piguet:1992yg,Piguet:1993ds} 
 can be understood as a direct consequence of the fact that the length
of the descent of the gauge anomaly is $4$, while the length of the
descent of all the other cohomological classes coupled to the action is $0$.
These considerations are purely cohomological, so that they do 
not depend on the way the gauge is fixed or on power counting
restrictions. Furthermore, this aproach to the 
Adler-Bardeen theorem does not require the use of the Callan-Symanzik
equation or assumptions on the beta functions of the theory.  

The article is organized as follows. In the next section, we review
the relevant features of the extended antifield formalism, with
special care devoted to the derivation of the
quantum BRST differential in the anomalous case. 
The characterization of local BRST cohomological groups
according to the lengths of their descents and their lifts is
explained in section \ref{char}.
The fourth section contains the main result on the lengths of the
descents and the lifts of the anomalies. An alternative derivation of
the main results clarifying the underlying mechanism is presented in
section \ref{alt}, where differentials controling
the one loop anomalies arising in the renormalization of 
BRST cohomological groups are
introduced. Anomalous Yang-Mills theory and the Adler-Bardeen theorem are
discussed in section \ref{AdBar}.

\section{The extended antifield formalism}
\label{sec1}

\subsection{Classical theory}

The Batalin-Vilkovisky formalism \cite{Batalin:1981jr,Batalin:1983jr} 
allows to formulate the BRST differential controling the 
gauge symmetries under renormalization for generic 
gauge theories. 
The formalism can be extended so as to 
include (non linear) global symmetries with a generic algebra 
\cite{Brandt:1998cz}. This 
is achieved by coupling the 
BRST cohomological classes in negative ghost numbers with constant
ghosts. There is a further extension to include the 
BRST cohomological classes in all the ghost numbers
\cite{Barnich:1999qz}, 
which allows to take into account in a systematic way all higher 
order cohomological constraints due to the antibracket maps 
\cite{Barnich:1998ph}. 

The completely extended formalism is obtained by 
first computing a basis for the local BRST cohomological classes
$H^{g,n}(s|d)$\footnote{We
  consider the local BRST cohomology in the space of Lorentz-invariant
  polynomials 
  or formal power seris in the $dx^\mu$, the coupling constants, 
the fields, antifields and their derivatives.}
associated to the standard differential $s$
and coupling those classes that are not already contained in the
solution of the standard master equation 
with the help of new independent coupling constants. 
This action can then be extended by terms of
higher orders in the new couplings in such a way that, if we denote by 
$\xi^A$ all the couplings corresponding to the independent 
BRST cohomological classes, the
resulting action $S$ satisfies the extended master equation
\be
\frac{1}{2}(S,S)+\Delta_c S=0.\label{1}
\ee
The BRST differential associated to
the solution of the extended master equation is 
\be
\bar s=(S,\cdot)+\Delta^L_c,
\ee
where $\D_c={\partial^R\cdot\over\partial\xi^A}f^A$, while 
$\Delta^L_c=(-)^Af^A{\partial^L\over\partial\xi^A}$, with $f^A$
depending at least quadratically on the couplings $\xi$ alone. Both
derivations satisfy 
$(\Delta_c)^2=0=(\Delta^L_c)^2$. 
Since there is no dependence on the fields and the
antifields, $\Delta^L_c(A,B)=(\Delta^L_cA,B)+(-)^{A+1}(A,\Delta^L_c
B)$, with the appropriate version holding for the right differential
$\Delta_c$. 

Besides the information on the invariance of the original
action under the non trivial gauge symmetries and their commutator algebra
contained in the standard solution of the master equation, the
extended master equation contains the information on the antibracket
algebra of all local BRST cohomological classes associated to the
standard BRST differential $s$. 
These classes contain the generators of all the 
generalized non trivial symmetries 
\footnote{Higher order symmetries have to be taken into
account in the generic case \cite{Brandt:1998cz}.} of the
theory in negative ghost number. This is the reason why the extended master
equation encodes in particular the invariance of the original action under all
the non trivial global symmetries as well as their commutator
algebra. In ghost number zero, the local BRST cohomological classes 
$H^{g,n}(s|d)$ contain
all the non trivial generalized observables of the 
theory\footnote{They correspond to local functionals built out of
  the original fields and their derivatives that are gauge invariant
  when the gauge covariant equations of motion hold, and that are linearily 
independent even when the gauge covariant equations of motion hold
(see e.g. \cite{Henneaux:1992ig}).}. In ghost number $1$, these
classes contain all the anomaly candidates that could affect the standard
Batalin-Vilkovisky master equation and in ghost number strictly
greater than $1$, they contain the anomaly candidates in ghost number
$g+1$ that could occur because the anomaly candidates in ghost number
$g$ have been coupled to the action. 

The cohomology of BRST differential $\bar s$ of the extended formalism 
in the space $F$ of $\xi$ dependent local 
functionals in the fields, the antifields and their derivatives, 
is isomorphic to the cohomology of 
\be
s_{\Delta_c}=[\Delta_c,\cdot]
\ee
in the space of graded right
derivations $\lambda=\frac{\partial^R\cdot}{\partial
  \xi^A}\lambda^A$, with $\lambda^A$ a function of $\xi$ 
alone, $[\cdot,\cdot]$ being the graded commutator for graded right
derivations. 
If $\mu$ is a $s_{\Delta_c}$ cocycle, the corresponding $\bar s$ cocycle is
given by $\mu S= \frac{\partial^R S}{\partial
  \xi^B}\mu^B$. In particular, the general solution to the system
of equations
\be
\left\lbrace 
\begin{array}{c}
\mu S+\bar s C  =0,\\
s_{\Delta_c}\mu  = 0,
\end{array} 
\right.
\label{8}
\ee
is given by
\be
\left\{\begin{array}{c}
C=\bar s D+ \nu S,\\
\mu=s_{\Delta_c}\nu.\end{array}\right.\label{9}
\ee

\subsection{Quantum theory}

In the usual version of the BRST-Zinn-Justin-Batalin-Vilkovisky set-up,
there are two main issues to be considered (see
e.g. \cite{Bonneau:1990xu,Piguet:1995er}): stability and anomalies.

\subsubsection{Stability}
  
The problem of stability is 
the question if to every local BRST cohomological class $H^{0,n}(s|d)$ 
in ghost number
$0$, there corresponds an independent coupling of the standard master
equation. The
extended formalism solves this problem by construction, because
all standard cohomological classes have been coupled with independent
couplings. Indeed, in the extended formalism, the differential
controling the ``instabilities'', i.e., the divergencies and/or
counterterms, is the differential $\bar s$, and according to the
previous section, $\bar s A=0$ implies $A=\mu S+\bar s B$, where $\mu$ 
belongs to $H(s_{\Delta_c})$, so that the non trivial part of $A$ can
indeed be absorbed by a modification of the couplings of the extended
master equation (see \cite{Barnich:1999qz} for more details).  

In different words, the extended antifield formalism guarantees
``renormalizability in the modern sense'' \cite{Gomis:1996jp} for all
gauge theories. 
Of course, it will be often convenient not to couple all 
the local BRST cohomological classes but only a subset needed to
guarantee that the theory is stable. 

\subsubsection{Anomalous Zinn-Justin equation}
\label{sec2}

In the standard set-up, the question of anomalies is mostly reduced to the
computation of the local BRST cohomological group $H^{1,n}(s|d)$ 
in ghost number $1$ and to a
discussion of the coefficients of the corresponding classes. In the
presence of anomalies, there is no differential 
on the quantum level associated to the anomalously broken Zinn-Justin
equation for the effective action. In the
extended antifield formalism however, because all the local BRST
cohomological classes in positive ghost numbers have been coupled to the
solution of the master equation, the broken Zinn-Justin equation can
be written as a functional differential equation and an 
associated differential exists, even in the presence of anomalies. To
show this, is the object of the remainder of this subsection
\footnote{We rederive section 4 of \cite{Barnich:1998ke} in a more
appropriate notation, 
insisting on the existence of the quantum BRST differential
in the anomalous case and its relation to its classical counterpart $\bar
s$. Note that the relation after (4.9) of that paper 
should read $s_Q \frac{\partial^R\cdot}{\partial
\xi^A}\sigma^A=\frac{1}{2}\frac{\partial^R\cdot}{\partial \xi^A}
[\sigma,\sigma]^A$ instead of $s_Q \frac{\partial^R\cdot}{\partial
\xi^A}\sigma^A=0$.} 

The quantum action principle
\cite{Lowenstein:1971jk,Lam:1972mb,Lam:1973qa} 
applied to (\ref{1}) gives
\be
\frac{1}{2}(\G,\G)+\D_c\G=\hbar{\cal A}\circ\G\label{2},
\ee
where $\G$ is the renormalized generating functional for 1PI vertices
associated to the solution $S$ of the extended master equation
and the local functional ${\cal A}$ is an element of 
$F$ in ghost number $1$. 
Applying
$(\G,\cdot)+\D^L_c$ to (\ref{2}), the l.h.s vanishes identically
because of the 
graded Jacobi identity for the antibracket and the properties of
$\D_c$, so that one gets
the consistency condition $(\G,{\cal A}\circ\G)+\D_c^L{\cal A}\circ\G
=0$. 
To lowest order
in $\hbar$, this gives $\bar s {\cal A}=0$, the general solution of
which can be writen as 
\be
{\cal A}=-{\partial^R S\over\partial \xi^A}\Delta^A_1+\bar s \Sigma_1,
\ee
with $[\Delta_c,\Delta_1]=0$, because of the relation between the
$\bar s$ and the $s_{\Delta_c}$ cohomologies discussed in the previous
subsection. 

If
one now defines ${S}^1=S-\hbar\Sigma_1$, the
corresponding generating functional admits the expansion 
$\G^1=\G-\hbar\Sigma_1+O(\hbar^2)$ and satisfies
$\frac{1}{2}(\G^1,\G^1)+\D^1\G^1=O(\hbar^2)$, where 
$\Delta^1=\Delta_c+\hbar 
\Delta_1$.  On the
other hand, the quantum action principle applied to
$\frac{1}{2}(S^1,S^1)+\D^1 S^1=O(\hbar)$ implies
$\frac{1}{2}(\G^1,\G^1)+\D^1\G^1=\hbar\bar{\cal A}\circ\G^1$, for a local
functional $\bar{\cal A}$. Comparing the two
expressions, we deduce that 
\be
\frac{1}{2}(\G^1,\G^1)+\D^1\G^1=\hbar^2{\cal A}^\prime\circ\G^1,
\ee
for a local functional ${\cal A}^\prime$.
Applying now
$(\G^1,\cdot)+(\Delta^1)^L$, one gets as consistency condition
\be
\frac{1}{2}[\D^1,\D^1]\G^1+\hbar^2((\G^1,{\cal A}^\prime\circ\G^1)
+\Delta^1{\cal A}^\prime\circ\G^1)=0,
\ee
giving to lowest order 
\be
1/2[\D_1,\D_1]S+\bar s 
{\cal A}^\prime=0.
\ee
Since $1/2[\D_1,\D_1]$ is 
a $s_{\Delta_c}$ cocycle because of the graded Jacobi identity for the graded 
commutator, equations (\ref{8}) and (\ref{9}) of the previous section
imply that the general solution to this equation is 
\be
\frac{1}{2}[\D_1,\D_1]+[\Delta_c,\D_2]=0, \\
{\cal A}^\prime=\bar s \Sigma_2-{\partial^R
  S\over\partial\xi^B}\D^B_2.
\ee 
The redefinition $S^2=S^1-\hbar^2\Sigma_2$ then allows to achieve 
\be
\frac{1}{2}(\G^2,\G^2)+\D^2\G^2=
\hbar^3{\cal A}^{\prime\prime}\circ\G^2,
\ee
for a local functional ${\cal A}^{\prime\prime}$, with 
$\D^2=\D^1+\hbar^2\D_2$.
The reasoning can be
pushed recursively to all orders with the result 
\be
\frac{1}{2}(\G^\infty,\G^\infty)+\D^\infty\G^\infty=0,\label{qsm}
\ee
where ${\G}^\infty$ is associated to the action 
${S}^\infty=S-\Sigma_{k=1}\hbar^k\Sigma_k$ and 
$\D^\infty=\D_c+\hbar \D_1 +\hbar^2 \D_2+\dots$ satisfies 
$(\D^\infty)^2=0$.
The associated quantum BRST differential is
\be
s^q=(\G^\infty,\cdot)+(\D^\infty)^L. 
\ee
In the limit $\hbar$ going to zero, it 
coincides with the classical differential $\bar s$. 

In the extended antifield formalism, the anomalous Zinn-Justin
equation can thus be written as a functional differential equation for the
renormalized effective action. The derivations $\D_1,\D_2,\dots$ are
guaranteed to exist due to the quantum action principles. They
satisfy a priori cohomological restrictions due to the fact that 
the differential $\D^\infty$ is a formal deformation with deformation
parameter $\hbar$ of the differential $\D_c$. In the context of chiral 
Yang-Mills theories, where $\Delta_c=0$, 
an anomalous master equation of the form
(\ref{qsm}) for the renormalized effective action has appeared for the 
first time in \cite{Costa:1977pd}. 

\subsubsection{Renormalization of local BRST cohomological classes 
of the extended formalism}

Associated quantum extension of the classical BRST cohomological classes 
$\lambda_0S$ of the extended formalism are given by $\lambda_0
  \Gamma^\infty$. By acting with
$\lambda_0$ on (\ref{qsm}),
one gets 
\be
s^q\lambda_0
  \Gamma^\infty=-[\Delta^\infty,\lambda_0]
\Gamma^\infty. 
\ee
The derivation $[\Delta^\infty,\lambda_0]=\sum_{n\geq L}\hbar^n \mu_n$ 
on the right hand side is of order at least $\hbar$, $L\geq 1$, because 
$[\Delta_c,\lambda_0]=0$ and corresponds to the anomaly in the invariant
renormalization of $\lambda_0
  S$. The question then is whether there
exists modified quantum extension $\lambda^\infty
  \Gamma^\infty$, with
${\lambda}^\infty=\lambda_0+\hbar\lambda_1+\dots$, such that 
\be
s^q\lambda^\infty
  \Gamma^\infty=0.
\ee
This is for instance the case if
$\lambda_0$ corresponds to the 
trivial cohomological class, $\lambda_0=[\Delta_c,\nu_0]$. The searched
for extension can then simply be taken to be
$\lambda^\infty=[\Delta^\infty,\nu_0]$. 

Because $[\Delta^\infty,[\Delta^\infty,\lambda_0]]=0$, 
the lowest order part of the anomaly, $\mu_L$ is an $s_{\Delta_c}$
cocyle, $[\Delta_c,\mu_L]=0$. Suppose that $\mu_L$ is a trivial
solution to this equation, $\mu_L=-[\Delta_c,\lambda_L]$. The modified
quantum extension $(\lambda_0+\hbar^L\lambda_L)\Gamma^\infty$
allows to push the anomaly to order $L+1$. 

Hence, to lowest order in $\hbar$, the non trivial
part of the anomaly in the renormalization of a classical cohomological 
class
$H^g(s_{\Delta_c})$ is constrained to belong to
$H^{g+1}(s_{\Delta_c})$. All the
quantum information on this anomaly is encoded in the derivations 
$\Delta_1,\Delta_2,\dots$ and the whole discussion 
has been shifted from local functionals to derivations of
functions of the coupling constants. 

\section{Characterization of local BRST cohomological classes}
\label{char}

In this and the following section, we decompose the space $\Omega$ of
Lorentz-invariant polynomials or formal powers series in the $dx^\mu$,
the couplings $\xi^A$, the fields, antifields and their derivatives
into the direct sum of the constants and the remaining part,
$\Omega={\bf R}\oplus \Omega_+$. We have $\bar s\alpha=0=d\alpha$,
for a constant $\alpha$ and
$d\Omega_+ \subset
\Omega_+$. We furthermore assume that if $\bar s \omega=\alpha$ 
for a constant $\alpha$, then $\alpha=0$, which amounts to assuming
that the equations of motions are consistent (see the discussion in
chapter 9 of \cite{BBHP}).  This means that 
$\bar s \tilde\Omega \subset \tilde\Omega$. Hence, we can consider
the cohomological groups $H(\bar s,\Omega_+)$, $H(\bar
s|d,\Omega_+)$ and $H(d,\Omega_+)$.

By analyzing the cohomological groups $H(\bar s|d)$ (the space
$\Omega_+$ being always understood in the following) using descent 
equations, one can prove \cite{Dubois-Violette:1985jb}
that the elements
of these groups can be classified into chains of length $r$ with an
obstruction to further lifts and chains of length $s$ 
whose lifts are unobstructed,
i.e., chains with a non trivial element in degree $n$ 
(see also \cite{Talon:1985dz,Brandt,Henneaux:1999rp};
we follow here the notations of the review \cite{BBHP}, 
where explicit proofs of the statements below can be found).
More precisely, we have 
$H^p(d)=0$, $p\leq n-1$, and there exists a basis 
\begin{eqnarray} 
\{[h^0_{i_r}],[\hat
h_{i_r}],[e^0_{\alpha_s}]\}\label{bass} 
\end{eqnarray} 
of $H(\bar s)$, for
$r=0,\dots,n-1$, $s=0,\dots,n$, such that a corresponding basis of
$H(\bar s|d)$ is
given by 
\begin{eqnarray}
\{[h^q_{i_r}],[e^p_{\alpha_s}]\}\label{bassd}
\end{eqnarray} 
for $q=0,\dots r$ and $p=0,\dots,s$, with 
\begin{eqnarray} 
\begin{array}{c}
\bar s h^{r+1}_{i_r}+d h^{r}_{i_r}=\hat h_{i_r},\\ 
\bar s h^r_{i_r}+ d
h^{r-1}_{i_r}=0,\\ 
{\vdots}\label{prop1}\\ 
\bar s h^1_{i_r}+ dh^0_{i_r}=0,\\ 
\bar s h^0_{i_r}=0, 
\end{array}
\end{eqnarray} 
and 
\begin{eqnarray} 
\begin{array}{c}
{\rm form\ degree}\ e^s_{\alpha_s}=n,\ de^s_{\alpha_s}=0, \\ 
\bar s e^{s}_{\alpha_s}+ d e^{s-1}_{\alpha_s} =0,\\ 
{\vdots}\label{prop2}\\
\bar s e^1_{\alpha_s}+ d e^0_{\alpha_s}=0,\\ 
\bar s
e^0_{\alpha_s}=0. 
\end{array}
\end{eqnarray} 
The cohomological group $H(\bar s)$ can thus be decomposed into elements 
$[e^0_{\alpha_s}]$ that are bottoms of unobstructed chains of length
$s$,
elements $[h^0_{i_r}]$ that are 
bottoms of obstructed chains of length $r$ and obstructions $[\hat
h_{i_r}]$ to chains of length $r$.
  
For the cohomological group $H(\bar s|d)$, 
the element $[h^l_{i_r}]$ is said to be the element of level $l$ of a
chain of length $r$ with obstruction; it has $l$ non trivial descents and 
$r-l$ non trivial lifts~; while the element
$[e^l_{\alpha_s}]$ is said to be the element of level $l$ of a chain of
length $s$ without obstructions, it has $l$ non trivial descents and
$s-l$ non trivial lifts.  

One can furthermore show that 
the general solution to a set of descent equations involving 
at most $l$ steps, 
$\bar s\omega^l+d\omega^{l-1}=0,$
$\bar s\omega^{l-1}+d\omega^{l-2}=0,\dots,$ $\bar s\omega^0=0$, 
can be written in terms of the above elements as   
\be
\omega^l=
\sum_{q=0}^l\sum_{r=l-q}^{n-1} \lambda^{i_r}_{q}h^{l-q}_{i_r}
+\sum_{p=0}^l\sum_{s=l-p}^n \mu^{\alpha_s}_{p}e^{l-p}_{\alpha_s}
+\bar s\eta^l +d[\eta^{l-1}+\sum_{r=0}^{n-1}\nu^{(l)i_r}h^r_{i_r}],
\label{lenl} 
\ee
with $\eta^{-1}=0$. This means that a $\bar s$ modulo $d$ cocycle
which has $l$ non trivial descents, is a linear combination
of all elements of the chains (\ref{prop1}) and (\ref{prop2}) which have 
$l$ or less non trivial descents. 
If such a linear combination is $\bar s$ modulo $d$ trivial,
the coefficients of the linear combination must vanish, i.e.,   
\be
\sum_{q=0}^{l}\sum_{r=l-q}^{n-1}
\lambda^{i_r}_{q}h^{l-q}_{i_r} +\sum_{p=0}^l\sum_{s= l-p}^n
\mu^{\alpha_s}_{p}e^{l-p}_{\alpha_s}=\bar s(\ )+d(\ ).
\label{cotr} 
\ee 
implies that $\lambda^{i_r}_{q}=0=\mu^{\alpha_s}_{p}$.

\section{Lengths of descent and lifts of lowest order anomalies}
\label{sec4}
Let us now investigate the anomalies in the BRST invariant 
renormalization of a chain (\ref{prop2}) of length $s$ without 
obstructions. We follow the approach of 
\cite{Lucchesi:1987sp}, which consists in considering simultaneously the
anomalies for a whole chain of descent equations
(for a review, see \cite{Piguet:1995er}). 
Using the quantum action principles, one can prove (as in lemma 1 in 
the appendix of \cite{Barnich:1998ke}) that if $a$ is a local form,
then 
\begin{eqnarray}
\bar s a\circ\G^\infty
=s^q a\circ\G^\infty+\hbar b\circ\G^\infty,
\end{eqnarray} 
for some local form $b$.
When applied to the chain (\ref{prop2}), we find that 
the quantum version of this chain is 
\begin{eqnarray}
\begin{array}{c}
{\rm form\ degree}\ e^s_{\alpha_s}\circ\G^\infty =n,\ 
de^s_{\alpha_s}\circ\G^\infty =0,\\ 
s^q e^{s}_{\alpha_s}\circ\G^\infty+ d e^{s-1}_{\alpha_s}\circ\G^\infty  =
\hbar a^s_{\alpha_s}\circ\G^\infty,\\ 
{\vdots}\label{q1}\\
s^q e^1_{\alpha_s}\circ\G^\infty+ d e^0_{\alpha_s}\circ\G^\infty  =
\hbar a^1_{\alpha_s}\circ\G^\infty,\\ 
s^q e^0_{\alpha_s}\circ\G^\infty  =\hbar a^0_{\alpha_s}\circ\G^\infty
, 
\end{array}
\end{eqnarray} 
where $a^l_{\alpha_s}\circ\G^\infty=a^s_{\alpha_s}+O(\hbar)$ for a local 
function $a^l_{\alpha_s}$. 
Applying $s^q$, we get the consistency condition,
\begin{eqnarray} 
\begin{array}{c}
{\rm form\ degree}\ a^s_{\alpha_s}\circ\G^\infty =n,\ 
da^s_{\alpha_s}\circ\G^\infty=0 \\ 
s^q a^{s}_{\alpha_s}\circ\G^\infty+ d a^{s-1}_{\alpha_s}\circ\G^\infty =
0,\\ 
{\vdots}\label{con1}\\
s^q a^1_{\alpha_s}\circ\G^\infty+ d a^0_{\alpha_s}\circ\G^\infty=
0,\\ 
s^q a^0_{\alpha_s}\circ\G^\infty=0.
\end{array}
\end{eqnarray} 
At lowest order in $\hbar$, we get 
\begin{eqnarray} 
\begin{array}{c}
{\rm form\ degree}\ a^s_{\alpha_s}=n,\ 
da^s_{\alpha_s}=0 \\ 
\bar s a^{s}_{\alpha_s}+ d a^{s-1}_{\alpha_s} =
0,\\ 
{\vdots}\label{locon1}\\
\bar s a^1_{\alpha_s}+ d a^0_{\alpha_s}=
0,\\ 
\bar s a^0_{\alpha_s}=0
.
\end{array} 
\end{eqnarray}
Using equation (\ref{lenl}) and the fact that the form
degree of $a^s_{\alpha_s}$ is $n$, it follows that 
\be
a^l_{\alpha_s}=
\sum_{p=0}^l
{\mu^{\beta_{s-p}}_{p}}_{\alpha_s}
e^{l-p}_{\beta_{s-p}}
+\bar s\eta^l_{\alpha_s} +d[\eta^{l-1}_{\alpha_s}+
\sum_{r\geq 0}\nu^{(l)i_r}_{\alpha_s}h^r_{i_r}],\label{res1}
\ee
for $l=0,\dots,s$.
This gives our first result:

{\em The anomaly in the renormalization of an element of level $l$ 
of a chain of length $s$ without obstructions 
involves at most elements of chains of the same type with less non
trivial descents and the same number of non trivial lifts.}

For the anomalies for a chain with obstruction (\ref{prop1}), we get, 
\begin{eqnarray} 
\begin{array}{c}
s^q h^{r+1}_{i_r}\circ\G^\infty+ 
dh^r_{i_r}\circ\G^\infty 
=\hat h_{i_r}\circ\G^\infty+\hbar a^{r+1}_{i_r}\circ\G^\infty,\\ 
s^q h^{r}_{i_r}\circ\G^\infty+ d h^{r-1}_{i_r}\circ\G^\infty  =
\hbar a^r_{i_r}\circ\G^\infty,\\ 
{\vdots} \label{q2}\\
s^q h^1_{i_r}\circ\G^\infty+ d h^0_{i_r}\circ\G^\infty  =
\hbar a^1_{i_r}\circ\G^\infty,\\ 
s^q h^0_{i_r}\circ\G^\infty  =\hbar a^0_{i_r}\circ\G^\infty
, 
\end{array}
\end{eqnarray} 
We also have 
\be
s^q \hat h_{i_r}\circ\G^\infty=-\hbar
\hat a_{i_r}\circ\G^\infty.\label{qo}
\ee
Applying $s^q$ gives 
$s^q \hat a_{i_r}\circ\G^\infty=0$ and then to lowest order, $\bar s
\hat a_{i_r}=0$. Applying now $s^q$ to the chain (\ref{q2}) gives
\be
\begin{array}{c}
s^q a^{r+1}_{i_r}\circ\G^\infty+ da^r_{i_r}\circ\G^\infty 
=\hat a_{i_r}\circ\G^\infty,\\ 
s^q a^{r}_{i_r}\circ\G^\infty+ d a^{r-1}_{i_r}\circ\G^\infty  =
0,\\ 
{\vdots} \label{con2}\\
s^q a^1_{i_r}\circ\G^\infty+ d a^1_{i_r}\circ\G^\infty  =
0,\\ 
s^q  a^0_{i_r}\circ\G^\infty  =0
,
\end{array} 
\ee
and to lowest order,
\be
\begin{array}{c}
\bar s a^{r+1}_{i_r}+ da^r_{i_r}
=\hat a_{i_r},\\ 
\bar s a^{r}_{i_r}+ d a^{r-1}_{i_r}  =
0,\\ 
{\vdots} \label{locon2}\\
\bar s a^1_{i_r}+ d a^1_{i_r}  =
0,\\ 
\bar s  a^0_{i_r}  =0
,
\end{array} 
\ee
On the one hand, it follows from (\ref{lenl}) that 
\be
a^l_{i_r}=\sum_{q=0}^l\sum_{r^\prime= r-q}^{n-1} 
{\lambda^{j_{r^\prime}}_{q}}_{i_{r}}h^{l-q}_{j_{r^\prime}}
+\sum_{p=0}^l\sum_{s = r-p+1}^n
{\mu^{\beta_{s}}_{p}}_{i_r}
e^{l-p}_{\beta_{s}}
+\bar s\eta^l_{i_r}+d[\eta^{l-1}_{i_r}+
\sum_{r^\prime= 0}^{n-1}\nu^{(l)j_{r^\prime}}_{i_r}
h^{r^\prime}_{j_{r^\prime}}],\label{res2}
\ee
for $l=0,\dots,r$,
while on the other hand, the cohomology of $\bar s$ implies
\be
\hat a_{i_r}=\sum_{r^\prime= 0}^{n-1}\alpha^{j_{r^\prime}}_{i_r}
\hat h_{j_{r^\prime}}+\sum_{r^\prime= 0}^{n-1}\beta^{j_{r^\prime}}_{i_r}
h^0_{i_r}+\sum_{s=0}^n\gamma^{\alpha_s}_{i_r}e^0_{\alpha_s}+\bar s
\hat o_{i_r}.\label{4.10}
\ee
Applying $d$ to (\ref{res2}) at $l=r$ gives
\be
d a^r_{i_r}=-\sum_{q=0}^{r}\sum_{r^\prime=r-q+1}^{n-1}
{\lambda^{i_r^\prime}_{q}}_{i_{r}}\bar s h^{r-q+1}_{i_{r^\prime}}
-\sum_{p=0}^{r}\sum_{s=r-p+1}^n 
{\mu^{\beta_{s}}_{p}}_{i_r}
\bar s e^{r-p+1}_{\beta_{s}}
-\bar sd \eta^r_{i_r}\nonumber\\+
\sum_{q=0}^{r}{\lambda^{i_{r-q}}_{q}}_{i_{r}}(
-\bar s h^{r-q+1}_{i_{r-q}}+\hat h_{i_{r-q}}).\label{4.11}
\ee
Injecting now (\ref{4.10}) and (\ref{4.11}) into the first equation of
(\ref{locon2})
gives first of all
$\beta^{j_{r^\prime}}_{i_r}=0=\gamma^{\alpha_s}_{i_r}$ and also
\be
\hat a_{i_r}=\sum_{q=0}^r{\lambda^{j_{r-q}}_{q}}_{i_r}
\hat h_{j_{r-q}}+\bar s
\hat o_{i_r},\label{res0}
\ee
and then,
\be
\bar s 
(a^{r+1}_{i_r}-\sum_{q=0}^{r}\sum_{r^\prime=r-q+1}^{n-1} 
{\lambda^{i_{r^\prime}}_{q}}_{i_{r}}h^{r-q+1}_{i_{r^\prime}}
-\sum_{p=0}^{r}\sum_{s=r-p+1}^n  
{\mu^{\beta_{s}}_{p}}_{i_r}
e^{r-p+1}_{\beta_{s}}\nonumber\\
-\sum_{q=0}^{r}{\lambda^{i_{r-q}}_{q}}_{i_r}
h^{r-q+1}_{i_{r-q}}-d \eta^r_{i_r}-\hat o_{i_r})=0,
\ee
so that, using the cohomology of $\bar s$,
\be
a^{r+1}_{i_r}=\sum_{q=0}^{r+1}\sum_{r^\prime=r-q}^{n-1} 
{\lambda^{i_{r^\prime}}_{q}}_{i_r}h^{r-q+1}_{i_{r^\prime}}
+\sum_{p=0}^{r+1}\sum_{s=r-p+1}^n 
{\mu^{\beta_{s}}_{p}}_{i_r}
e^{r-p+1}_{\beta_{s}}\nonumber\\
+d [\eta^r_{i_r}+\sum_{r^\prime=0}^{n-1}
\nu^{(r+1)j_{r^\prime}}_{i_r}h^{r^\prime}_{j_{r^\prime}}]
+\hat o_{i_r}+\bar s
\eta^{r+1}_{i_r}.\label{c5}
\ee
Our second result is then: 

{\em The anomaly in the renormalization of an element of level $l$ 
of a chain of length $r$ with obstructions 
involves at most elements of chains with obstructions with less non
trivial descents and more non trivial lifts and elements of 
chains without obstructions with less non trivial descents and 
strictly more non trivial lifts.}

Let us now rewrite (\ref{res1}) and (\ref{res2}) at $l$=0 as
\be
a^0_{\alpha_s}=
{\mu^{\beta_{s}}_{0}}_{\alpha_s}
e^{0}_{\beta_{s}}
+\bar s[\eta^0_{\alpha_s} -\sum_{r= 0}^{n-1}\nu^{(0)i_r}_{\alpha_s}
h^{r+1}_{i_r}]+
\sum_{r= 0}^{n-1}\nu^{(0)i_r}_{\alpha_s}\hat h_{i_r},\label{res7}\\
a^0_{i_r}=\sum_{r^\prime= r}^{n-1}
{\lambda^{j_{r^\prime}}_{0}}_{i_{r}}h^{0}_{j_{r^\prime}}
+\sum_{s = r+1}^{n}
{\mu^{\beta_{s}}_{0}}_{i_r}
e^{0}_{\beta_{s}}
+\bar s[\eta^0_{i_r}-\sum_{r^\prime= 0}^{n-1}\nu^{(0)j_{r^\prime}}_{i_r}
h^{r^\prime+1}_{j_{r^\prime}}] +
\sum_{r^\prime=0}^{n-1}\nu^{(0)j_{r^\prime}}_{i_r}
\hat h_{j_{r^\prime}}.\label{res8}
\ee 
Combined with (\ref{res0}), our third result on anomalies 
in the renormalization of elements of $H(\bar s)$ is accordingly:

{\em The anomaly in the renormalization of obstructions to chains of
  length $r$ 
  involves at most obstructions to shorther chains; the anomaly in 
  the renormalization of bottoms of unobstructed chains of length $s$
  involves at most bottoms of unobstructed chains of the same length
  and obstructions to chains of all possible lengths; the anomaly in
  the renormalization of bottoms of obstructed chains of length $r$
  involves at most bottoms of obstructed chains of 
greater length, bottoms of unobstructed chains of strictly 
greater length and obstructions to chains of all possible lengths.}

\section{Differentials associated to one loop anomalies}
\label{alt}

A different, more compact, way to formulate and prove
the results of section \ref{char} and \ref{sec4} is to use the 
exact couple describing the descent 
and the associated spectral sequence \cite{Dubois-Violette:1985jb}. 

Indeed, the diagram  
\begin{eqnarray}
\begin{array}{ccc}
H(\bar s|d )\stackrel{{\cal D}}{\longrightarrow}
H(\bar s|d)\\
i_0\nwarrow\ \swarrow l_0\\
H(\bar s)
\end{array}\label{ec}
\end{eqnarray}
can be shown to be exact at all corners. 
The various maps are defined as follows:  $i_0$ is the map which
consists in regarding an element of $H(\bar s)$ as an element of
$H(\bar s|d)$,  
$i_0:H(\bar s)\longrightarrow
H(\bar s|d)$, with $i_0[a]=[a]$. It is well defined  
because every $\bar s$ cocycle is a 
$\bar s$ cocycle
modulo $d$ and every $\bar s$ coboundary is a $\bar s$ 
coboundary modulo $d$. 
The descent homomorphism 
${\cal D}:
H^{k,l}(\bar s|d)\longrightarrow
H^{k+1,l-1}(\bar s|d)$ with 
${\cal D}[a]=[b]$, if $\bar s a + db=0$ is 
well defined because of the triviality of the cohomology of $d$ in
form degree $p\leq n-1$.   
Finally, the map $l_0:H^{k+1,l-1}(\bar s|d)\longrightarrow
H^{k+1,l}(\bar s)$ is defined by $l_0[a]=[d a]$. It is well defined
because the relation $\{\bar s ,d\}=0$ implies that
it maps
cocycles to cocycles and coboundaries to coboundaries. 

Associated to such an exact
couple $(H(\bar s|d),K_0=H(\bar s))$, 
one can associate in a standard
way derived exact couples $({\cal D}^rH(\bar s|d), K_r)$,
\begin{eqnarray}
\begin{array}{ccc}
{\cal D}^r H(\bar s|d )\stackrel{{\cal D}}{\longrightarrow}
{\cal D}^r H(\bar s|d)\\
i_r\nwarrow\ \swarrow l_r\\
K_r,
\end{array}\label{ecr}
\end{eqnarray}
and a spectral sequence $K_{r+1}=H(d_r,K_r)$, with 
$K_0\equiv H(\bar s)$. The maps of these exact couples are
defined recursively as follows: the map $d_{r-1}=l_{r-1}\circ i_{r-1}$ 
can be shown to be a differential, the map 
$i_r$ is the map induced by $i_{r-1}$ in $K_r$, while $l_r {\cal D} 
[a]=l_{r-1}[a]$.

Explicitly, 
the differential $d_0: K_0\longrightarrow K_0$ is defined by
$d_0[a]=[da]$, where $\bar s a=0$.  
An element $k_r\in K_r$ is
identified with the equivalence class $[a]_r$ of an element $[a]\in 
H(\bar s)$, 
where
$[a]\sim_r [a^\prime]$ if $[a]-[a^\prime]\in \oplus_{q=0}^{r-1}{\rm
  im}
\ d_q$. 
The relations $d_q k_{r}=0$, $q=0,\dots,r-1$ mean
that $k_r$ is a bottom that can be lifted at least $r$ times, i.e., 
there exist
$c_1,\dots,c_{r+1}$ such that $\bar s a=0, da+\bar s c_1=0,\dots, 
dc_{r-1}+\bar s c_{r}=0$. Then, the differential $d_r$ is defined by 
$d_r k_r=[dc_r]_r$.

Because there are no forms
of form degree higher than $n$, ${\cal D}^{n+1} H(\bar
s|d)=0$ and $d_n\equiv 0$ so that the construction stops at $r=n$. 

The space of local forms $\Omega$ is decomposed as 
$\Omega=E_0\oplus G\oplus \bar s G \oplus {\bf R}$, with  
$E_0\simeq K_0=H(\bar s)$. 
If we define $E_r,F_{r-1}\subset E_{r-1}$ through 
$E_{r-1}=E_{r-1}\oplus E_r\oplus 
d_{r-1} F_{r-1}$ with $E_r\simeq K_r$, we get the decomposition
\be
E_0=F_0\oplus \dots\oplus F_{n-1}\oplus E_n\oplus
d_{r-1} F_{r-1}\oplus\dots\oplus d_0 F_0. \label{decomp}
\ee
The $e^0_{\alpha_s}$ are elements of a basis of $E_n$
that can be lifted $s$ times before hitting form degree $n$, i.e., 
that are of form degree $n-s$, while 
$\hat h_{i_r}$ and $h^0_{i_r}$ are elements of a basis
of $d_r F_r $ and $F_r$ respectively. This sums up the results of 
section \ref{char}. 

Let us now define the linear map 
\be
\delta_0: H^g(\bar s)\longrightarrow
H^{g+1}(\bar s),\\
\delta_0[a]=[b],\ \mbox{where} 
\ s^q a\circ
\G^\infty=\hbar b\circ\G^\infty. 
\ee
The map associates to a given BRST
cohomological class the non trivial order $\hbar$, i.e., 1 loop
contribution of its anomaly.

The map is well defined, because the 
consistency condition implies that $\bar s b=0$, and if $a=\bar s c$,
$a\circ \G^\infty =s^q c\circ \G^\infty+\hbar d\circ \G^\infty$, so that 
$s^q a\circ \G^\infty=\hbar s^q d\circ \G^\infty$, meaning that $b=\bar
s d$, so that the map does not depend on the choice of the
representative. Furthermore, this map is a differential
\be
\delta_0^2=0. 
\ee 
Indeed, if $[a]=\delta_0[c]$, we have $a\circ\G^\infty 
=\frac{1}{\hbar}s^q c\circ \G^\infty$. It follows that $s^q
a\circ\G^\infty=0$. 
A BRST cohomological class which is a  $\delta_0$-cocycle has no
1-loop anomaly, while a BRST cohomological class which is a 
 $\delta_0$-coboundary is the 1-loop anomaly of some other 
BRST cohomological class. 

We thus have two differentials in $K_0=H(\bar
s)$, $d_0$ introduced above and $\delta_0$. These 
differentials anticommute, 
\be
\{d_0,\delta_0\}=0. 
\ee
Indeed, if $s^q a\circ
\G^\infty=\hbar b\circ\G^\infty$, $d_0\delta_0 [a]=d_0[b]=[db]$, while 
$\delta_0 d_0[a]=\delta_0 [da]$, and $s^q(da\circ\G^\infty)=s^q 
d(a\circ \G^\infty +\hbar c\circ\G^\infty)=-d s^q (a
\circ \G^\infty +\hbar c\circ\G^\infty)=-\hbar d(b\circ \G^\infty+
s^q c\circ\G^\infty)$, so that $\delta_0 [da]=[-d(b+\bar s
c)]=-[db]$. 

The relation $d_0 \delta_0 [a]= -\delta_0 d_0[a]$ means:
\begin{itemize}
\item if $[a]$ belongs to ${\rm im}\ d_0$, $[a]=d_0[b]$, then  
$\delta_0[a]=-d_0\delta_0[b]$,
i.e., if $[a]$ represents an obstruction to the lift of an element
$[b]$, its anomaly represents minus the obstruction to the lift of the
anomaly of $[b]$, 
\item if $d_0 [a]=0$, then $d_0\delta_0[a]=0$, i.e., if $[a]$ can be
  lifted, then so does its anomaly $\delta_0[a]$,
\item the
  anomaly in a bottom $[a]$ of $K_0$ that cannot be lifted is minus 
the anomaly of the corresponding obstruction, up to elements that can
  be lifted.
\end{itemize}
If we organize the space $E_0\simeq K_0=H(\bar s)$ as $E_0=F_0\oplus 
E_1\oplus d_0 F_0$, with $E_1\simeq K_1$, 
we have shown that 1-loop contribution
to the anomaly of an element in one of these subspaces belongs to the
same subspace or to a subspace that stands to the right. Together with
the last point of the previous list, this sums up the results for the 
elements of $H(\bar s)$, i.e., for the obstructions and the 
bottoms contained in (\ref{res0}), (\ref{res7}) and (\ref{res8}) 
restricted to $r=0$. 

In the same way, these results for all $r$ and $s$ 
follow from the fact that $\delta_0$ induces a well-defined
differential (also called $\delta_0$ in the following) in the spaces
$K_r$, anticommuting with $d_r$, $\delta_0:
K_r\longrightarrow K_r$ with $\{\delta_0,d_r\}=0$. 

Indeed, suppose
this result to be true for $K_0,\dots, K_{r-1}$, $d_0,\dots,d_{r-1}$. 
An element $[a]_r\in K_r$ 
satisfies $\bar s a=0$, $da +\bar s c_1=0$, $\dots,$ $dc_{r-1}+\bar s
c_r=0$. This implies $s^q a\circ\G^\infty=\hbar b\circ\G^\infty$, 
$d a\circ\G^\infty +s^q
c_1\circ\G^\infty=\hbar f_1\circ\G^\infty$, $\dots,$ 
$dc_{r-1}\circ\G^\infty+s^q
c_r\circ\G^\infty=\hbar f_r\circ\G^\infty$. Applying $s^q$ gives to
lowest order 
 $\bar s b=0$, $db +\bar s f_1=0$, $\dots,$ $df_{r-1}+\bar s
f_r=0$, so that $[b]_r$ is well defined. Suppose now that 
$[a]_r=d_0[g_0]_0+\dots+d_{r-1}[g_{r-1}]_{r-1}$. Anticommutativity of
$\delta_0$ with $d_0,\dots, d_{r-1}$ then implies that
$\delta_0[a]_r=0$. Hence, $\delta_0$ does not depend on the
representative and is well defined in $K_r$. Finally,
$d_r\delta_0[a]_r=d_r[b]_r=[df_r]_r$,
while $\delta_0d_r[a]_r=\delta_0 [dc_r]_r$, and $s^q
(dc_r)\circ\G^\infty=s^qd(c_r\circ \G^\infty+\hbar c^\prime\circ\G^\infty)=
-d( s^q c_r\circ\G^\infty+\hbar s^q c^\prime\circ \G^\infty)$, so that 
$\delta_0 [dc_r]_r=-[d(f_r+\bar s c^\prime)]_r=-d_r\delta_0[a]_r$. 

The results (\ref{res0}), (\ref{res7}) and (\ref{res8}) can then be 
summarized by the statement that an anomaly in one of the subspaces
of the decomposition (\ref{decomp})
must belong to the same subspace or to one that stands to the right; 
furthermore, the part of the anomaly of an element of $F_i$ in $F_i$
is minus the part of the anomaly of
the corresponding element of $d_i F_i$ in $d_{i}F_{i}$.

In order to recover the results for elements of $H(\bar s|d)$, 
we define $\Delta_0$ to be the equivalent of $\delta_0$ for modulo $d$
BRST cohomological classes, 
$\Delta_0 [a]=[b]$, for $[a],[b]\in H(\bar s|d)$, 
where $\bar s a+dm=0$, $\bar s m+du=0$, 
$s^qa\circ \G^\infty+d(m\circ\G^\infty)=\hbar b\circ \G^\infty$, 
$s^qm\circ \G^\infty+d(u\circ\G^\infty)=\hbar n\circ \G^\infty$.
Indeed, the map is well defined because 
the consistency condition implies to lowest order
$\bar s b+ dn=0$, while if $a=\bar s c+ dg$, we have $m=\bar s g +d
u$, so that $s^q a\circ\G^\infty+dm\circ\G^\infty=\hbar b\circ
\G^\infty$ gives $s^q(s^q c\circ\G^\infty+d g\circ\G^\infty+\hbar
f\circ\G^\infty)+d(s^q g\circ\G^\infty+du\circ\G^\infty+\hbar
v\circ\G^\infty)=\hbar b\circ\G^\infty$, which implies $b=\bar s f+dv$ as it
should. 

The following properties are straightforward to check: 
$[\Delta_0,{\cal D}]=0$,
$l \Delta_0 =-\delta_0 l$, $i_0\delta_0=\Delta_0 i_0$. 
One says (see e.g. \cite{Hu}, Chapter VIII.9) that
$(\Delta_0,\delta_0)$ is a mapping of the exact couple $(H(\bar
s|d),H(\bar s))$. 

The previous result, that $\delta_0$ induces well
defined maps in the spaces of the spectral sequence anticommuting with 
the differentials $d_r$, follows directly from the way the spectral 
sequence is associated to an exact couple. The relation between
(\ref{res1}), (\ref{res2}) at $l=0$ and (\ref{res7}), (\ref{res8}) is
summarized by $i_0\delta_0=\Delta_0 i_0$~; the relations between 
(\ref{res1}), (\ref{res2}) at different values of $l$ are summarized
by $[\Delta_0,{\cal D}]=0$~; finally, the relation between
(\ref{res2}) at $l=r$ and (\ref{res0}) is summarized by 
$l \Delta_0 =-\delta_0 l$. 

Note that in this case, we have furthermore the property that
$\Delta_0$ is a differential, $\Delta_0^2=0$.

{\bf Remark:} It follows from the above analysis that the relevant
property of the differentials $d_r$ is $\{\delta_0,d_r\}=0$. 
This means that analogous results that constrain the anomalies to
belong to particular subspaces of $H(\bar s)$ or $H(\bar s|d)$ 
can be derived if one can find maps $\lambda_0:H(\bar s)\longrightarrow 
H(\bar s)$, respectively $\Lambda_0:H(\bar s|d)\longrightarrow 
H(\bar s|d)$ such that $[\delta_0,\lambda_0]=0$, respectively
$[\Delta_0,\Lambda_0]=0$. 

\section{Adler-Bardeen theorem revisited}
\label{AdBar}

We now apply the ideas of the extended antifield formalism to standard
Yang-Mills theory. In this case, it is sufficient for our purpose to
couple the
local BRST cohomology classes in ghost number $0$ and ghost number $1$,
because this will be enough, under some assumptions stated explicitly below,  
to guarantee stability and to control the
anomalies. The starting point action contains from the beginning 
a coupling to the Adler-Bardeen
anomaly as in 
\cite{Costa:1977pd,Aoyama:1981yw,Tonin:1992wf}, 
with additional couplings to possibly
higher dimensional gauge invariant operators, 
if one does not want to restrict oneself to the
power counting renormalizable case \cite{Gomis:1996jp}. 
More precicley, the starting point 
is the action 
\be
S_\rho=\int d^4x\ [-\frac{1}{4g^2}F^{\mu\nu}_I F_{\mu\nu}^I+
L^{\rm kin}_{\rm matter}(\psi^i,D_\mu\psi^i)]\nonumber \\
+\int d^4x\ [-D_\mu C^IA^{*\mu}_I+ C^I
T^i_{Ij}\psi^j\psi^*_i
-\frac{1}{2}C^IC^J{f_{JI}}^KC^*_K]\nonumber  \\
+g^i \int d^4x\ {\cal O}_i +{\cal A}\rho,\label{6.1}
\ee
satisfying the master equation
\be
\frac{1}{2}(S_\rho,S_\rho)=0.
\ee
The Lagrangian $L^{\rm kin}_{\rm matter}(\psi^i,D_\mu\psi^i)$ is the gauge 
invariant extension of the kinetic terms for the matter fields $\psi^i$. 
For simplicity, we assume the gauge group to be $SU(3)$. The 
${\cal O}_i$ are gauge invariant local functions built 
out of the field strengths $F_{\mu\nu}^I$, the matter fields $\psi^i$ 
and their covariant derivatives such 
that the $\int d^4x\ {\cal O}_i$ (which can, but need not, be assumed to 
be power counting renormalizable) and $\int d^4x\
-\frac{1}{4g^2}F^{\mu\nu}_I F_{\mu\nu}^I$ are linearily independent
even when the gauge covariant equations of motions hold. Finally,
${\cal A}=\int {\rm Tr}\ [Cd(AdA+\frac{1}{2}A^3)]$ 
is the Adler-Bardeen gauge anomaly, $g$ is the gauge coupling
constant, $g^i$ are coupling constants for the other gauge invariant
operators, while $\rho$ is a Grassmann odd coupling constant 
with ghost number $-1$ for the 
Adler-Bardeen anomaly.
In this particular case, 
$\Delta_c=0$. This can be traced back to the fact that all
representatives of the local BRST cohomological classes in ghost
number $0$ and $1$ can be choosen to be independent of the
antifields. 

The gauge is fixed by introducing the cohomologically 
trivial non minimal
sector consisting of the antighost $\bar C^I$ and the Lagrange
multiplier $B^I$ and their antifields. One adds to the action (\ref{6.1}) 
the term 
$\int d^4x\ \bar C_I^* B^I$ and chooses an appropriate gauge 
fixing fermion $\Psi$, 
which is used to  
generate an anticanonical transformation in the fields and
antifields such that the propagators of the
theory are well defined. The gauge fixing is irrelevant for the 
cohomological considerations below. 

For the question of stability and anomalies, we have to analyze the
cohomology $H^{0,4}(s_\rho|d)$ and $H^{1,4}(s_\rho|d)$ in the space of 
functions in the couplings 
$g,g^i,\rho$ with coefficients that are Lorentz invariant polynomials
in the $dx^\mu$, the fields, the antifields and their derivatives. 

In order to compute this cohomology, we decompose, as in 
\cite{Costa:1977pd},
 both the BRST differential $s_\rho$ and the local forms into 
parts independent of
$\rho$ and parts linear in $\rho$. Explicitly, $s_\rho=s_0+s_1$, where 
$s$ is the standard BRST differential associated to the solution
$S_{\rho=0}$ of the master equation, while $s_1=({\cal A}\rho,\cdot)$.
The $\rho$ independent part of the cocycle condition 
$s_\rho \omega(\rho)+d\eta(\rho)=0$ 
in form degree $4$ gives (see e.g. \cite{BBHP} for a review) 
$\omega(0)=\alpha(g,g^i)d^4x\ F^{\mu\nu}_I
F_{\mu\nu}^I+\alpha^j(g,g^i)d^4x\ {\cal O}_i +s(\ ) +d(\ )$. This
implies $\omega(\rho)=\alpha(g,g^i)d^4x\ F^{\mu\nu}_I
F_{\mu\nu}^I+\alpha^j(g,g^i)d^4x\ {\cal
  O}_i+\omega^\prime_1\rho +s_\rho(\ )+d(\ ).$ Because $d^4x\ F^{\mu\nu}_I
F_{\mu\nu}^I$ and $d^4x\ {\cal O}_i$ are also $s_1$ closed and $\rho^2=0$, the
cocycle condition reduces to $s\omega^\prime_1+d(\ )=0$, where the 
ghost number of $\omega^\prime_1$ is $1$. It follows that  $\omega^\prime_1=
\lambda(g,g^i){\rm Tr}\ [Cd(AdA+\frac{1}{2}A^3)]+s(\ )+d(\
)$. Hence, the general solution of the consistency condition in the
space of local functionals in ghost number $0$ is given by 
\be
\alpha(g,g^i)\frac{\partial S_\rho}{\partial g}+
\alpha^j(g,g^i)\frac{\partial S_\rho}{\partial g^i}+ 
\frac{\partial^R S_\rho}{\partial
  \rho}\rho\lambda(g,g^i)+(S_\rho,\Xi_\rho)
\label{6.4}
\ee 
for some local functional $\Xi_\rho$ in ghost number $-1$. 
This implies that the theory is stable. 

Similarily, in ghost number $1$, the $\rho$ independent part of the
cohomology gives as only anomaly candidate the Adler-Bardeen anomaly.
There could however be a $\rho$ linear non trivial contribution to the anomaly,
because the cohomology of 
$s$ in form degree $4$ and ghost number $2$ is not necessarily empty 
(see \cite{BBHP}, section 12.4), contrary to the 
claim in \cite{Costa:1977pd}. 
More precisely, to each $x^\mu$-independent, 
gauge and Lorentz invariant non trivial conserved 
current $j_\Delta=j^\mu_\Delta\epsilon_{\mu\nu_1\nu_2\nu_3}
dx^{\nu_1}dx^{\nu_2}dx^{\nu_3}$, there corresponds the cohomological class
$V^{2,4}_\Delta=j_\Delta [{\rm Tr} C^3]^1+{\it antifield\ dependent\ terms}$
in $H^{2,n}(s|d)$, with $s[{\rm Tr} C^3]^1+{\rm Tr} C^3=0$.(There 
could in principle be another type of antifield dependent cohomology classes 
in exceptional situations \cite{BBHP}, which we exclude 
from the present considerations).

If there are such non trivial currents $j_\Delta$, we have to change our 
starting point and also couple the ``anomaly for anomaly candidates''
$V^{2,4}$ 
from the beginning with couplings in ghost number $-2$. But then the 
cohomology of $s$ in ghost number $3$ becomes relevant. 
There are plenty of such classes, for instance classes of the form 
$d^4x\ I\ {\rm Tr}C^3$, where $I$ are invariant functions built 
out of the field strengths, the matter fields and their covaraint 
derivatives. In this way, one is led to use the full extended 
antifield formalism as described in section \ref{sec1} 
with all BRST cohomology 
classes in positive ghost 
number coupled from the beginning. In the case where the algebra of the 
non trivial symmetries associated to the currents $j_\Delta$ is non abelian, 
the operator $\Delta_c$ will be non vanishing at the classical
level and involve in particular the structure constants of this algebra.

Another possibility is to try to show that the
anomaly candidates $V^{2,4}_\Delta$ do not effectively arise in the theory, 
by using higher order cohomological restrictions: as in the proof of the
absence of similiar instabilities in the presence of abelian factors
for standard Yang-Mills theories (see \cite{Barnich:1999qz}, Appendix A)
one couples with external fields gauge invariant functions that break
the symmetries associated to the currents $j_\Delta$. In this way, one 
eliminates the currents $j_\Delta$ and the associated anomaly for
anomaly candidates $V^{2,4}_\Delta$ from the extended theory. At the
end of the computations, the external fields can be put to zero. 

Because this discussion is not central to the argument below, we will simply
assume here that 
the observables ${\cal O}_i$ are
such that there are no non trivial currents $j_\Delta$ and thus no
anomaly for anomaly candidates $V^{2,4}_{\Delta}$ in the
theory. 
The general solution of the consistency condition in ghost number $1$
is then given by 
\be
\frac{\partial S_\rho}{\partial \rho}\sigma(g,g^i)
+(S_\rho,\Sigma_\rho), \label{6.5}
\ee
for some local functional $\Sigma_\rho$ in ghost number
$0$.

By standard arguments, using in addition the same reasoning as in 
section \ref{sec2}, it follows from (\ref{6.4}) and (\ref{6.5}) 
that the model is renormalizable and 
that the renormalized 
generating functional for 1 particle irreducible vertex functions
$\Gamma_\rho$ satisfies 
\be
\frac{1}{2}(\Gamma_\rho,\Gamma_\rho)+\frac{\partial^R\Gamma_\rho}{\partial 
  \rho}\hbar \sigma(g,g^i)=0,\label{fun}
\ee
where $\sigma(g,g^i)$ is a formal power series in $\hbar$. Hence, in
this case $\Delta^\infty=\frac{\partial^R\cdot}{\partial 
  \rho}\hbar \sigma(g,g^i)$ and the quantum BRST differential is 
$s^q=(\Gamma_\rho,\cdot)-\hbar\sigma(g,g^i){\partial^L}{\partial 
  \rho}$. (It can be shown along standard lines that the coefficient 
$\sigma(g,g^i)$ does not depend on the parameters of the gauge fixing.)  

Let us now investigate the renormalization of the operators 
$d^4x\ F^{\mu\nu}_I F_{\mu\nu}^I$ and $d^4x\ {\cal O}_i$. According to 
the classification in section \ref{char}, they are both of the type 
$e^0_{\alpha_0}$, because they are non trivial
bottoms in maximal form degree $4$, while the Adler-Bardeen anomaly 
${\rm Tr}\ [Cd(AdA+\frac{1}{2}A^3)]$ is of the type
$e^4_{\alpha_4}$, as it descends to the non trivial bottom 
${\rm Tr}C^5$. Because there is no $e^0_{\alpha_0}$ in ghost
number $1$ (and form degree 4) 
and no $h^r_{i_r}$ in form degree $3$ and ghost number $1$,
equation (\ref{res1}) implies that the lowest order contribution to the 
anomaly in the renormalization of $d^4x\ F^{\mu\nu}_I F_{\mu\nu}^I$
and of $d^4x\ {\cal O}_i$ is $s_\rho$ exact and can thus be absorbed
through a BRST breaking counterterm added to these operators. This
reasoning can be pushed to all orders, with the result that one can
achieve, through the addition of suitable counterterms,
\be
s^q [d^4x\ F^{\mu\nu}_I F_{\mu\nu}^I\circ\G_\rho]=0,\label{36}\\
s^q [d^4x\ {\cal O}_i\circ\G_\rho]=0\label{37}.
\ee
If we now apply $-\frac{g}{2}\frac{\partial}{\partial g}$,
respectively $\frac{\partial}{\partial g^j}$ to (\ref{fun}), we get 
on the one hand,
\be
s^q [\int d^4x\ -\frac{1}{4g^2}F^{\mu\nu}_I F_{\mu\nu}^I\circ\G_\rho]+
\frac{\partial^R\Gamma_\rho}{\partial 
  \rho}\hbar [-\frac{g}{2}\frac{\partial\sigma(g,g^i)}{\partial g}]=0,\\
s^q [\int d^4x\ {\cal O}_i\circ\G_\rho]+\frac{\partial^R\Gamma_\rho}{\partial 
  \rho}\hbar [\frac{\partial\sigma(g,g^i)}{\partial g^j}]=0.
\ee 
Comparing on the other hand with the integrated versions of (\ref{36})
and (\ref{37}), we deduce that 
\be
\frac{\partial\sigma(g,g^i)}{\partial g}=0,\ 
\frac{\partial\sigma(g,g^i)}{\partial g^j}=0,
\ee
which is an expression of the Adler-Bardeen theorem (see e.g. section
6.3.2 of \cite{Piguet:1995er}).

\section{Conclusion and outlook}

In the completely extended antifield formalism, where in particular 
potential anomaly candidates are coupled to the starting point action 
with independent couplings, the whole program of algebraic
renormalization can be extended to the case of theories with
gauge anomalies. For instance, it is possible to write the 
Callan-Symanzik equation to all orders as a functional
differential equation as in \cite{Barnich:1998ke}, by using
$\Delta^\infty$ instead of $\Delta_c$. The $\beta$ functions can then
be considered as anomaly coefficients in the renormalization
of the cohomology class corresponding to the generator of
dilatations. Vanishing results for $\beta$ functions
associated to couplings of cohomological classes with a non
trivial descent can then be obtained as a particular case of the
discussion of section \ref{sec4}. 

The discussion in section \ref{sec4} on the length of descents and 
lifts of BRST
cohomological classes and their anomalies does not rely on the use 
of the extended antifield formalism. It can be done along the same
lines in the 
standard set-up as long as one assumes the quantum theory to be  
anomaly free and stable, so that the standard Zinn-Justin equation
$\frac{1}{2}(\Gamma,\Gamma)=0$ holds.  
A virtue of the completely extended formalism is that it allows in
principle to treat the
general case.

\acknowledgments

This work has been partly supported by the ``Actions de
Recherche Concert{\'e}es" of the ``Direction de la Recherche
Scientifique - Communaut{\'e} Fran{\c c}aise de Belgique", by
IISN - Belgium (convention 4.4505.86). The author wants to thank
F.~Brandt and M.~Henneaux for useful discussions.

\bibliographystyle{unsrt}

\bibliography{master1}

\end{document}